# Rotation Curve of Galaxies by the Force Induced by Mass of Moving Particles

Kyuwook Ihm[1,*] and Kyoung-Jea Lee [2]

[1]Beamline research division, Pohang Accelerator Laboratory, Pohang, Kyungbuk 790-784, Korea

[2]Department of physics, POSTECH, Pohang, Kyungbuk 790-784, Korea

We suggest that there is a novel force which is generated by the mass of relatively moving particles. The new force which we named Mirinae Force is a counterpart of the magnetic force operating between electrically charged moving particles. Instead of using the conventional dark matter, we applied the mirinae force to a particular model system of the spiral galaxy in which most of the galaxy's mass is located within the central region where some portion of the inner mass is in revolving motion at a relativistic speed. The calculation yielded three important results that illustrate the existence of mirinae force and validate the proposed model: First, the mirinae force in this model explains why most of the matters in the galactic disk are in the circular motion which is similar to cycloid. Second, the mirinae force well explains not only the flat rotation curve but also the varied slope of the rotation curve observed in the spiral galaxies. Third, at the flat velocity of 220 Km/s, the inner mass of the Milky Way calculated by using the proposed model is $6.0 \times 10^{11}$ $M_\odot$, which is very close to $5.5 \times 10^{11}$ $M_\odot$ ($r$ <50 Kpc, including Leo I) estimated by using the latest kinematic information. This means that the mirinae force well takes the place of the dark matter of the Milky Way.

[*]To whom correspondence should be addressed. E-mail: johnet97@postech.ac.kr

General relativity is a theory which has been considered to describe experimental results related to gravity and cosmology most successfully to date. However, there are indications that the theory is incomplete because, for example, the problem of quantum gravity and the question of the reality of spacetime singularities remain open [1]. The rotation curve of galaxies is one of the questions under debate. There have been many attempts to explain the flat rotation curve of the galaxies regardless of distance which is in a discrepancy with the Newtonian prediction [2, 3]. Many studies have tried to find a correct form of the gravitational law or the distribution of hidden dark matter that fits well with observed data. Although these efforts have been rather phenomenologically successful, two important questions still remain: first, do we correctly take all the factors affecting the rotation curve of the galaxy into account? Second, do we have a correct theory of gravity?

In this study we suggest a novel force generated between relatively moving particles by their mass as a counterpart of the magnetic force between the electrically charged particles that are in motion, and we named it "mirinae force". If we pay attention to the fact that magnetic field is induced by a variation of electric field in a spacetime without the magnetic monopole, the expectation of alternative force field generated by a variation of gravitational field in the identical spacetime looks reasonable regardless of what is mass. Physical phenomena related to mass and charge have always been described in a separated path way probably because interactions between mass and charge have never been observed. However, mass and charge are not completely independent from each other. Unless the mass of the test particle is known, one cannot determine the amount of electric charge by measuring the acceleration of a test particle driven by Coulomb force generated by another particle with known value of $q$.

Mirinae force is derived by defining a single complex value called the grand-mass whose unit is kg. The grand-mass is assumed to characterize all the physical phenomena of particles related to both mass and electric charge. Grand-mass is defined as

$$M=m+i\alpha q \text{ [kg]} \tag{1}$$

where $m$ and $q$ are conventional mass and charge of a stationary particle in a inertial frame, and $\alpha$ is matching constant with an unit of kg/C. We first demonstrate that the grand-mass of the particles well describes the Newtonian mechanics including gravitational and Coulomb interactions and the momentum-energy conservation of special relativity. However, the description of the force between moving charged particles using grand-mass causes a novel force generated by mass between the moving particles as a counterpart of the magnetic force. This force is what we call mirinae force. We will show that the mirinae force is so small that its effects can be important only in a system of astronomical scale. If a huge mass flow exists near the galactic center, the mirinae force field would be produced inducing a circular motion of matters in the galactic disk, which is similar to a cycloid. This model based on the concept of mirinae force explains well the flat rotation curve found in a spiral galaxy, such as the Milky Way and M31, and it gives an orbit velocity and inner mass well accorded with recent estimation.

With the grand-mass the momentum which is conserved in all inertial reference frames can be written as

$$\vec{P}_G = \gamma M \frac{d\vec{r}}{dt} = \gamma (m \frac{d\vec{r}}{dt} + i\alpha q \frac{d\vec{r}}{dt}) \tag{2}$$

where, $\gamma = (1-v^2/c^2)^{-1/2}$ is Lorentz factor of a particle with velocity $v$. The energy-momentum conservation law still holds true. If all mass terms are replaced by the grand-mass, $M$, the result is reduced to $E^2 -(pc)^2=(mc^2)^2$.

The grand-force is defined as

$$\vec{F}_G = \frac{d\vec{P}_G}{dt} = m(\gamma \frac{d^2\vec{r}}{dt^2} + \frac{d\vec{r}}{dt}\frac{d\gamma}{dt}) + i\alpha q(\gamma \frac{d^2\vec{r}}{dt^2} + \frac{d\vec{r}}{dt}\frac{d\gamma}{dt}) = \vec{F}_m + i\vec{F}_q \tag{3}$$

In the expression of the grand-force, the nontrivial term is only mass related real part of $F_m$. This is identical to the fact that in a system with an isolated single particle with mass, $m$, and charge, $q$, the observer can measure only the inertial mass $m$ (Fig. 2a). The observer cannot notice the existence of electric charge included in a single particle system. This situation is well described by the imaginary representation of charge related term in (3). The measureable physical value of an isolated particle cannot be related with charge, $q$, because the charge related force term is in the imaginary part.

The grand-force acting on a stationary ($\gamma = 1$) particle $M_1$ located at $\vec{r}_1$ due to another particle $M_2$ at $\vec{r}_2$ can be expressed as, (Fig. 2b),

$$\vec{F}_{G_{M1}} = M_1 \frac{d^2 r}{dt^2} = G \frac{M_1 M_2}{|\vec{r}_{21}|^2} \hat{r}_{21} = \frac{G}{|\vec{r}_{21}|^2}(m_1 + i\alpha q_1)(m_2 + i\alpha q_2)\hat{r}_{21}$$
$$= \{G\frac{m_1 m_2}{|\vec{r}_{21}|^2} - G\alpha^2 \frac{q_1 q_2}{|\vec{r}_{21}|^2}\}\hat{r}_{21} + i\frac{G\alpha}{|\vec{r}_{21}|^2}(m_1 q_2 + m_2 q_1)\hat{r}_{21} = \vec{F}_{m_1} + i\vec{F}_{q_1} \tag{4}$$

where $G$ is the gravitational constant of $6.67\times10^{-11}$ [m$^3$/s$^2$Kg]. The constant $G\alpha^2$ shown in the second term of real force part should be equal to $1/4\pi\varepsilon_0=8.99\times10^9$ [m$^3$Kg/s$^2$C$^2$] when compared with that of Coulomb force. This gives $\alpha=1.161\times10^{10}$ [Kg/C], which show that the unit of the grand-mass in (1) is Kg. The directions of the two force terms in the real value part of (4) well accord with that of gravitational and Coulomb forces, i.e., attractive and repulsive. Interestingly, the charge related term is involved with the mass related force term, $\vec{F}_{m_1}$ in (4).

When there is a flow of particles with mass and charge, one can define the grand-mass current density as

$$\vec{J}_M = \{\rho_m(\vec{r},t) + i\alpha\rho_q(\vec{r},t)\}\vec{v}(\vec{r},t) = \vec{J}_m + i\alpha\vec{J}_q \tag{5}$$

where $\rho_m$, $\rho_q$ and $v$ are the density of mass, charge and the average drift velocity of particles at point $r$ at time $t$. If identical particles with density $n$ are in the flow, then the grand-mass current density is

$$\vec{J}_M = mn\vec{v} + i\alpha qn\vec{v} = \vec{J}_m + i\alpha\vec{J}_q \tag{6}$$

The magnetic force can be described by the grand-mass current density of moving particles. To do so, the electric current in the magnetic force law need to be replaced by grand-mass current, and constants of the resulting equation should be determined. The simplest example of Ampère's law is the force per unit length between two straight parallel current carrying conductors, which is expressed as (7) with magnetic constant $K$.

$$F_A = 2K\frac{I_1 I_2}{r}, \quad K = \frac{\mu_0}{4\pi} \tag{7}$$

Considering the identical system with two parallel straight lines carrying grand-mass current $I_{M1}$ and $I_{M2}$, and replacing magnetic constant $K$ with undefined one, $K_{GM}$, then we have

$$\begin{aligned} F_{I_M} &= 2K_{GM}\frac{I_{M_1}I_{M_2}}{r} = 2K_{GM}\frac{1}{r}(I_{m_1} + i\alpha I_{q_1})(I_{m_2} + i\alpha I_{q_2}) \\ &= 2K_{GM}\frac{I_{m_1}I_{m_2}}{r} - 2K_{GM}\alpha^2\frac{I_{q_1}I_{q_2}}{r} + i\alpha 2K_{GM}\frac{(I_{m_1}I_{q_2} + I_{m_2}I_{q_1})}{r} \end{aligned} \tag{8}$$

Comparing second term in (8) with Ampère's force in (7), one can have $K_{GM} = -\mu_0/4\pi\alpha^2$. Following the way that the magnetic constant $K$ in (7) determines the magnitude of the magnetic field generated by electric current in the Biot-Savart law, the force field generated by a grand-mass current density can be found using $K_{GM}$ as

$$\begin{aligned} \vec{B}_G &= K_{GM}\int\frac{\vec{J}_M \times \hat{r}}{r^2}dxdydz = K_{GM}\int\frac{(\vec{J}_m + i\alpha\vec{J}_q) \times \hat{r}}{r^2}d^3x \\ &= -\frac{\mu_0}{4\pi\alpha^2}\int\frac{\vec{J}_m \times \hat{r}}{r^2}d^3x - \frac{i}{\alpha}\frac{\mu_0}{4\pi}\int\frac{\vec{J}_q \times \hat{r}}{r^2}d^3x \\ &= \vec{B}_m + \frac{1}{i\alpha}\vec{B}_{B-S} \equiv \vec{B}_m + \frac{1}{i\alpha}\vec{B}_q \end{aligned} \tag{9}$$

$$\vec{B}_m \equiv -\frac{\mu_0}{4\pi\alpha^2}\int\frac{\vec{J}_m \times \hat{r}}{r^2}d^3x \;,\; \vec{B}_q \equiv \frac{\mu_0}{4\pi}\int\frac{\vec{J}_q \times \hat{r}}{r^2}d^3x = \vec{B}_{B-S} \tag{10}$$

The equation (9) has two terms. The real value term, $\vec{B}_m$, is induced by the mass flow and named mirinae field. The imaginary second term is induced by the electric current, and $\vec{B}_q(=\vec{B}_{B-S})$ represents Biot-Savart law. Note that the direction of the mirinae field is opposite to that of the magnetic field generated by positively charged moving particles. Now, the Lorentz force can be replaced by grand-Lorentz force expressed by the grand-mass as

$$\vec{F}_{G-L} = M(\vec{v} \times \vec{B}_G) = (m + i\alpha q)\vec{v} \times (\vec{B}_m + \frac{1}{i\alpha}\vec{B}_q)$$
$$= m(\vec{v} \times \vec{B}_m) + q(\vec{v} \times \vec{B}_q) + i\alpha\{q(\vec{v} \times \vec{B}_m) - \frac{1}{\alpha^2}m(\vec{v} \times \vec{B}_q)\} \quad (11)$$

In the eq. (11) $B_G$ is replaced by (9). The grand-Lorentz force has two real terms in which the first real term is mirinae force and the second real term is the Lorentz force. Imaginary term corresponds to the imaginary force appeared in (3).

To test the magnitude of the mirinae force compared to the Lorentz force, consider a grand-Lorentz force acting on a moving electron due to the grand-force field produced by flowing electrons. In this case the ratio of charge to mass is constant, i.e., $\Sigma = e/m_e$. Then, the mass current density of electrons is (see the first term in (6))

$$\vec{J}_{m_e} = m_e n\vec{v} = \frac{1}{\Sigma}en\vec{v} = -\frac{1}{\Sigma}\vec{J}_{q_e} \quad (12)$$

From (10) and (12) the mirinae field is

$$\vec{B}_{m_e} \equiv -\frac{\mu_0}{4\pi\alpha^2}\int\frac{\vec{J}_{m_e} \times \hat{r}}{r^2}d^3x = \frac{1}{\Sigma\alpha^2}\frac{\mu_0}{4\pi\alpha^2}\int\frac{\vec{J}_{q_e} \times \hat{r}}{r^2}d^3x = \frac{m_e}{e\alpha^2}\vec{B}_{q_e} \quad (13)$$

Then, the real term of grand-Lorentz force on a moving electron with velocity, $v$, is from (11) and (13)

$$\text{Re}(\vec{F}_{G-L}) = m_e(\vec{v} \times \vec{B}_{m_e}) - e(\vec{v} \times \vec{B}_{q_e}) = \{(\frac{m_e}{e\alpha})^2 - 1\}e(\vec{v} \times \vec{B}_{q_e}) \approx -e(\vec{v} \times \vec{B}_{q_e}) \quad (14)$$

This shows that the grand-Lorenz force on a moving electron is almost the same as the Lorentz force. The magnitude of mirinae force is $(m_e/e\alpha)^2 = 2.39 \times 10^{-43}$ of Lorentz force. This shows why the mirinae force had not been found in the laboratories system on earth until now.

However, for further discussion the mirinae force has to show some persuasive and meaningful physical examples. Due to such a small magnitude of the mirinae force when compared with Lorentz force, meaningful effects of the mirinae force should be sought among the systems in astronomical scale.

It is known that 70% of galaxies exhibit a variety of spiral structures. Among them galaxies with a central bulge in their disk have a rotation curve which is flat from centre to edge [3, 4]. However, it was expected that these galaxies would have a rotation curve that slopes down as the distance increases in the inverse square root relationship, in the same way as other systems with most of their mass in the centre, such as the solar system of planets. The solution of this discrepancy has been sought by introducing dark matter or modified Newtonian dynamics (MOND) because kinetics in most the galactic region except for the vicinity of the black hole can be described by the non-relativistic gravitational theory [5-7]. Unfortunately, these methods have not been always successful in describing the shape of the rotation curves and have not shown a deeper physical understanding [8].

We propose a novel model in which dark matter is excluded whereas most of the mass ($m_C$) is assumed to be near the galaxy center and some portion of them revolves at relativistic velocity to the same revolving direction as that of matter on the galactic disk. To simplify the calculation, the circular mass flow around the center of galaxy is considered as a circular loop of radius $a$ carrying the particle flow, lying in the x-y plane, and centered at the origin (Fig. 2). If the net electric charge is zero, the grand-mass current is $I_M = I_m (+ i\alpha I_q = 0)$. The current density has only one component in the $\phi$ direction, which can be expressed using delta function as

$$\vec{J}_m = J_{m,\phi}\hat{\phi} = I_m \sin\theta' \delta(\cos\theta') \frac{\delta(r'-a)}{a} \hat{\phi} \tag{15}$$

From equation (9) the force field generated by grand-mass flow has only mass term (mirinae field)

$$\vec{B}_G = \vec{B}_m = -\frac{\mu_0}{4\pi\alpha^2} \int \frac{\vec{J}_m \times \hat{r}}{r^2} d^3x \tag{16}$$

The solution can be easily found following the technique shown in the current loop problem solved in classical electrodynamics [9]. For rough estimation, assuming the case of $r \gg a$, the solution is found as (17). The mirinae field on the galactic disk, corresponding to x-y plane in Fig. 2, can be obtained from (17) at $\theta = \pi/2$ as (18).

$$\vec{B}_m = -\frac{\mu_0}{4\pi\alpha^2}(I_m \pi a^2)\frac{\sin\theta}{r^3}\hat{\theta} - \frac{\mu_0}{2\pi\alpha^2}(I_m \pi a^2)\frac{\cos\theta}{r^3}\hat{r} \tag{17}$$

$$\vec{B}_m(\theta = \frac{\pi}{2}) = -\frac{\mu_0}{4\pi\alpha^2}(I_m \pi a^2)\frac{1}{r^3}\hat{\theta} \equiv -\frac{C}{r^3}\hat{\theta} \tag{18}$$

where the constant $C \equiv \mu_0 I_m \pi a^2 / 4\pi\alpha^2$ [m$^3$/s] in (18). Now, the grand-Lorentz force acting on the moving particle $M' = m' (+ i\alpha q' = 0)$ on the galactic disk at distance $r$ with velocity $\vec{v}$ is

$$\vec{F}_{G-L} = m'(\vec{v} \times \vec{B}_m) = -m'\frac{C}{r^3}(\vec{v} \times \hat{\theta}) \tag{19}$$

If a particle $M'$ is moving with the velocity, $-\vec{v}\hat{r}$, as shown in figure 2a, the mirinae force acts on it toward $\hat{\phi}$ direction as indicated in (19), inducing a circular motion of particle $M'$ with the same revolving direction of as that of the mass current in a loop. It is notable that, in case of the Lorentz force a charged particle, $M''(=m''+i\alpha q'')$, with velocity $-\vec{v}\hat{r}$ is forced to move in the opposite revolving direction to that of electric current of the ring.

Consider another case where a particle $M'(=m')$ is in circular motion around the galaxy center with velocity $\vec{v}\hat{\phi}$ at distance $r$ (Fig. 2b). The particle $M'$ is attracted toward $-\hat{r}$ direction by the gravitational force induced by a huge centered grand-mass $M_C(=m_C)$, and repulsed toward $\hat{r}$ direction by the centrifugal force and the mirinae force due to $-(\vec{v}\times\hat{\theta})=v\hat{r}$ in (19). Assuming that the orbital motion of particle $M'$ is stable, i.e., $dr/dt\approx 0$, we have the relation written as

$$G\frac{m_C m'}{r^2} = m'\frac{v^2}{r} + m'v|B_m| \tag{20}$$

The orbit velocity of $M'$ satisfying the relation (20) can be obtained with (18) as

$$v = \frac{1}{2}\{\sqrt{(\frac{C}{r^2})^2 + \frac{4Gm_C}{r}} - \frac{C}{r^2}\} \tag{21}$$

If there is no mass flow ($C=0$), the velocity of (21) describes the rotation curve of the planets in the solar system according to Newtonian mechanics. Figure 3 shows the rotation curves of (21) for the cases that $C = 1\times 10^{44} \sim 1\times 10^{49}$ when the inner mass is $m_C = 5\times 10^{11}$ $M_\odot$. The rotation curve at $C \sim 1.0\times 10^{47}$ (m$^3$/s) shows a nearly flat velocity in the wide distance ranges and has a flat velocity value of 220~250 Km/s similar to that of the Milky Way and M31 [10-12]. However, as the orbit distance decreases below 7 Kpc, the velocities start to be out of the permissible range of the observed values. This is originated from over simplified calculation model in Figure 2. In the real galactic system the gravitational field as well as mirinae field varies significantly depending on the distance, especially in the region of the galactic bulge, because the inner mass and mass flow are spatially distributed [4, 12]. Therefore, detailed calculation of the rotation curve near the galactic center must take into account the spatial distribution of mass and mass flow. Variation of $C$ value by a factor of 10 from $C=1\times 10^{47}$ (m$^3$/s) makes the curve profile completely different, indicating that the profile of rotation curve is sensitively dependent on the $C$ value. The curve profile can be estimated by the slope of rotation curve at a specific distance, $r$, as a function of $C$ value, written as

$$\frac{dv}{dr} = \frac{C}{r^3}[1 - \frac{1 + Gm_C r^3/C^2}{\sqrt{1 + 4Gm_C r^3/C^2}}] \qquad (22)$$

To show the detailed dependence of the curve profile on the $C$, $dv/dr$ at $r = 15$ Kpc and $m_C = 5 \times 10^{11}$ M$_\odot$ is plotted as a function of $C$ using filled squares in Figure 4. The corresponding velocities from (21) are shown by circles in Figure 4. The orbit velocity of stars when the rotation curve is flat, i.e., flat velocity, can be obtained using the specific $C$ value at which $dv/dr$ is zero. As shown in Figure 4, the $C$ value causing the flat rotation curve is $C \approx 6.4 \times 10^{46}$ m$^3$/s at which the orbital velocity corresponds to 250 Km/s. The $C$ values above $1.0 \times 10^{49}$ m$^3$/s also induce the slope to be zero, but the region of those values can be excluded from our discussion because in this region the orbit velocity is also zero, which is not the case of the spiral galaxy. It is interesting note that the curve slope has both the negative and positive values in the vicinity of $C(dv/dr=0) = 6.4 \times 10^{46}$ m$^3$/s, which accords well with the observation that showed various slopes of rotation curve in the spiral galaxies [7].

The mirinae field in the form of $-C/r^3$ of (18) is derived from the model as described in Figure 2. The magnitude of $C(=\mu_0 I_m \pi a^2/4\pi\alpha^2)$ represents the strength of the mirinae field at specific distance of orbit. This model assumed that some portion of inner mass is in a circular motion with relativistic velocity, which is simplified as a circular loop carrying grad-mass with radius $a$. Mass current $I_m$ in $C$ is given by $\lambda_\gamma v$, where $\lambda_\gamma$ (= $\gamma \times$Mass of loop/$2\pi a$) is relativistic line density of mass which is in revolving motion at mean velocity $v$ along the loop. Assuming that 1% of $m_C$, $\sim 5.0 \times 10^9$ M$_\odot$ (= mass of the loop), is in the circular motion at relativistic speed at $\gamma = 1.88 \times 10^7$, and the radius of loop is $a = 100$ pc, then we have $C = 6.41 \times 10^{46}$ [m$^3$/s] at which $dv/dr = 0$ as shown in Figure 4. In case of the electron in relativistic motion the kinetic energy at $\gamma = 1.88 \times 10^7$ corresponds to about 9 TeV. Note that the loop is imaginary system which represents all possible mass flow generating mirinae field near the galactic center including the black hole which is generally believed to be a supermassive one. However, due to the lack of reliable information on the distribution of the inner mass and mass flow near the galactic center the detailed discussion on the relevance of the parameters of the loop could not be made here.

In Figures 3 and 4 we assumed $m_C = 5 \times 10^{11}$ M$_\odot$ and showed that the magnitude of $C$ and orbit velocity at specific distance can be obtained by the additional condition $dv/dr = 0$. The $C$ value that satisfy the condition of $dv/dr = 0$ in (22) and the corresponding orbit velocity are expressed as

$$C(dv/dr = 0, r = 15 Kpc) = \sqrt{\frac{Gm_C (15 Kpc)^3}{2}}$$

$$v(dv/dr = 0, r = 15 Kpc) = \sqrt{\frac{Gm_C}{2 \times 15 Kpc}} \qquad (23)$$

Figure 5 shows the dependence of $v(m_c, r=15$ Kpc$)$ and $C(m_c, r=15$ Kpc$)$ on the inner-mass, $m_C$, when $dv/dr=0$. In most of the spiral galaxies, the flat orbit velocities of stars fall in the range of 150 ~ 300 Km/s [7]. Within this range the inner mass can be estimated by (23), and, as shown in Figure 5, interestingly it is in the range of $1.5 \times 10^{11}$~$7 \times 10^{11}$ $M_\odot$, which is very similar to the estimated results using the kinematic information for the spiral galaxies without taking account of dark matter [10, 12, 13]. For example, at orbit velocity of 220 Km/s, the inner mass is estimated to be $6.0 \times 10^{11} M_\odot$ from Figure 5 or (23); and the orbit velocity of the Milky Way is about 220 Km/s and the mass ($r <50$ Kpc, including Leo I) is estimated to be $5.5 \times 10^{11}$ $M_\odot$ by using the latest kinematic information [12]. This result means that the mirinae force well takes the place of the dark matter of the Milky Way.

We suggested the existence of a mirinae force field generated by the mass current density similar to the magnetic force. When the concept of grand-mass is developed further in theory, one can obtain an equation on the gravitational-mirinae field which is a counterpart of the electromagnetic field derived from the Maxwell equation. We expect that the enormous mass flow in circular motion near the galaxy center would emit gravitational-mirinae wave at a speed of light analogous to the electromagnetic wave emitted from the accelerated charged particles.

In summary, we proposed that a novel force, named mirinae force, is operating between relatively moving particles by their mass. This force is derived as a counterpart of the magnetic force by introducing a grand-mass defined as $M=m+i\alpha q$ [kg]. To show the relevance of the grand-mass in the physical system, we proposed a model of a spiral galaxy in which some portion of the matter in the massive galactic center is revolving at a relativistic speed. The mirinae force in this model explains why most of the matters in the galactic disk are in the circular motion in a manner similar to the cycloid. Calculations in this model explain the various curve profiles of the orbit velocity reported by other researchers, including the flat rotation curve in the spiral galaxies. The inner mass of the the Milky Way estimated by this model gives $6.0 \times 10^{11}$ $M_\odot$ which is very close to $5.5 \times 10^{11}$ $M_\odot$ ($r <50$ Kpc, including Leo I) estimated by using the latest kinematic information.

**Acknowledgements.**

This work was supported by Basic Science Research Program through the National Research Foundation of Korea (NRF) funded by the Ministry of Education, Science and Technology (Grant No. 2011-0026550).

Figure 1. **Particles with grand-mass** (a) The force exerted on a particle with grand-mass $M$ in the complex space. In the system composed of an isolated single particle it is possible to measure only the inertial mass $m$ from the real term of grand-force. (b) Interactions between two particles with grand-mass $M_1$ and $M_2$.

Figure 2. **The proposed model for the spiral galaxy.** Most of the mass, $M_C=m_C$, of galaxy is near the galactic center, and some portion of the inner mass is in revolving motion, which is replaced by the loop carrying the grand-mass at relativistic speed. In this model the direction of the mirinae field is $-\hat{\theta}$ on the x-y plane to which the galactic disk belongs. (a) If the velocity of the star is $-v\hat{r}$, then the star is forced toward $\hat{\phi}$ direction, resulting in a circular motion. (b) If the velocity of the star is $v\hat{\phi}$, then the star feels the repulsive mirinae force toward $\hat{r}$ direction, as well as centrifugal force.

Figure 3. Rotation curves acquired from (21) when $m_C = 5\times10^{11}$ M$_\odot$ and $C = 1.0\times10^{44} \sim 1.0\times10^{49}$ [m$^3$/s]. $C$ value represents the strength of the mirinae force at distance $r$ produced by the mass carrying loop with radius $a$.

Figure 4. Slopes($dv/dr$) of rotation curves at $r$ =15 Kpc when $m_C = 5\times10^{11}$ M$_\odot$ as a function of $C$ values. In the $C$ range of $2.0\times10^{45} \sim 2.0\times10^{47}$ [m$^3$/s] the slopes of the rotation curve are changed from negative to positive values, showing that the varied curve profiles are made in this range. At Slopes($dv/dr$)=0, i.e., the flat rotation curve, the flat velocity is 250 Km/s.

Figure 5. Flat velocities and $C$ values as a function of the inner mass, $m_C$, when the condition of $dv/dr$ = 0 is satisfied at $r$=15 Kpc. In this condition, if the flat velocity is given, the inner mass, $m_C$, and $C$ value are exclusively determined. The inner mass of the spiral galaxies with flat velocity in the range of 150~300 Km/s are estimated to be in the ranges of $1.5\times10^{11} \sim 6.5\times10^{11}$ M$_\odot$.

Figure 1

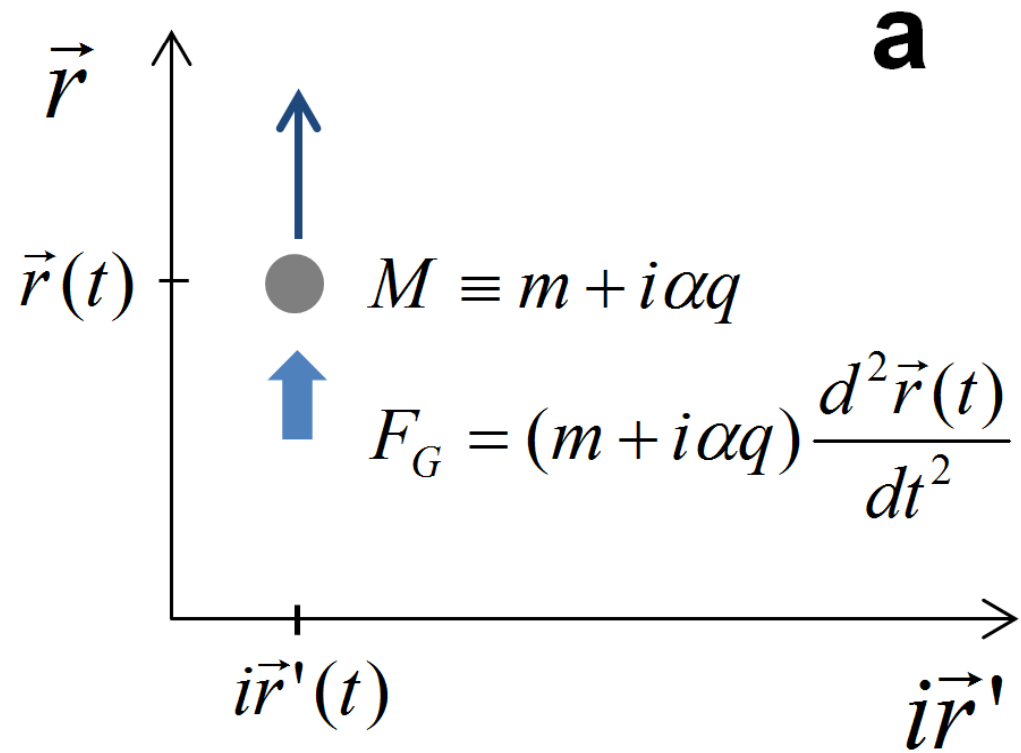

Figure 1

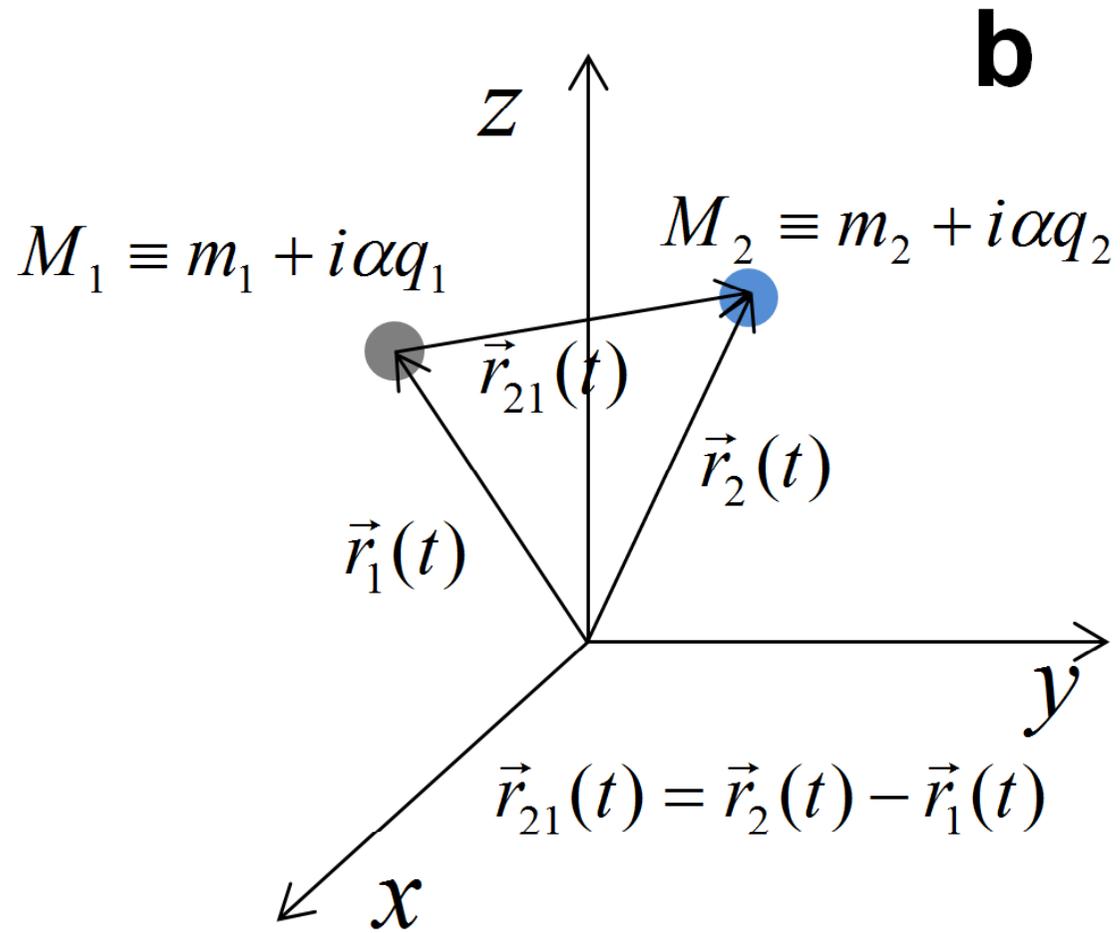

Figure 2

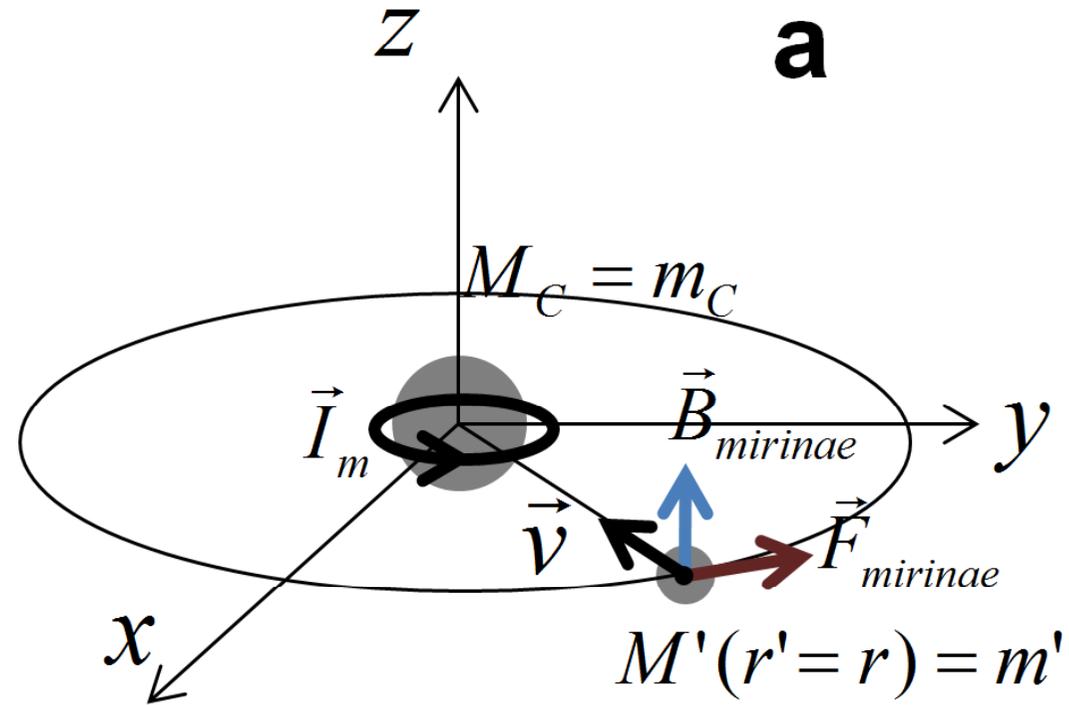

Figure 2

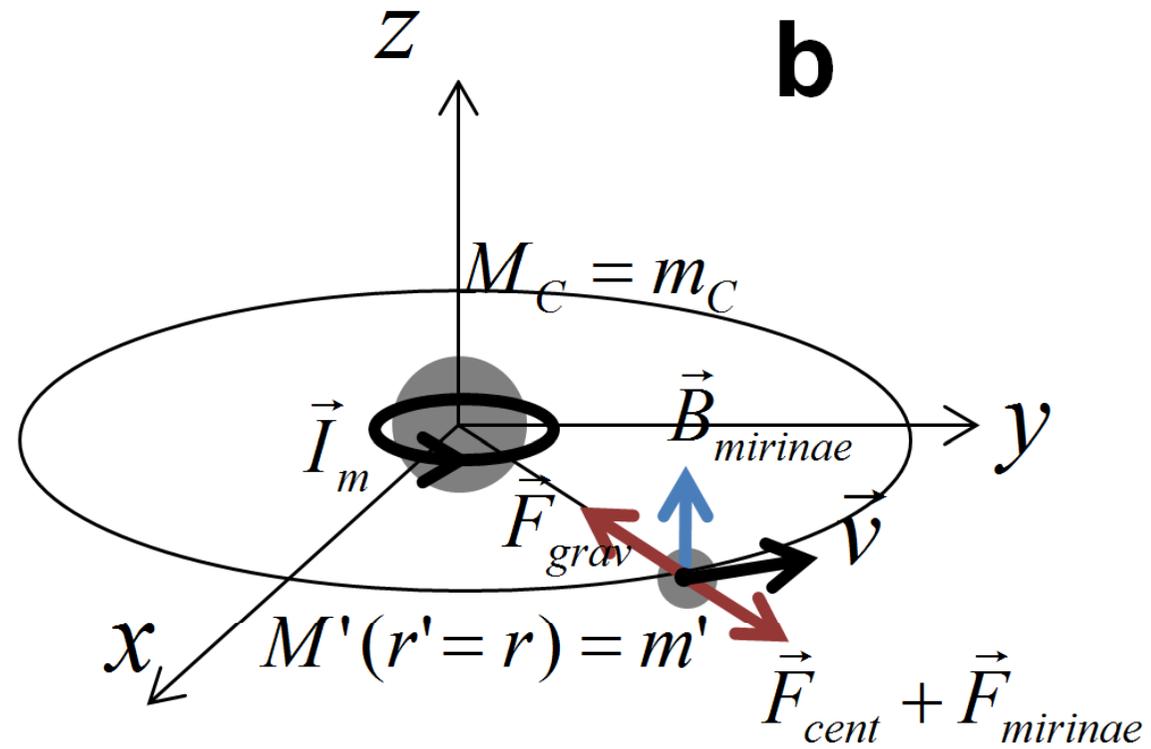

Figure 3

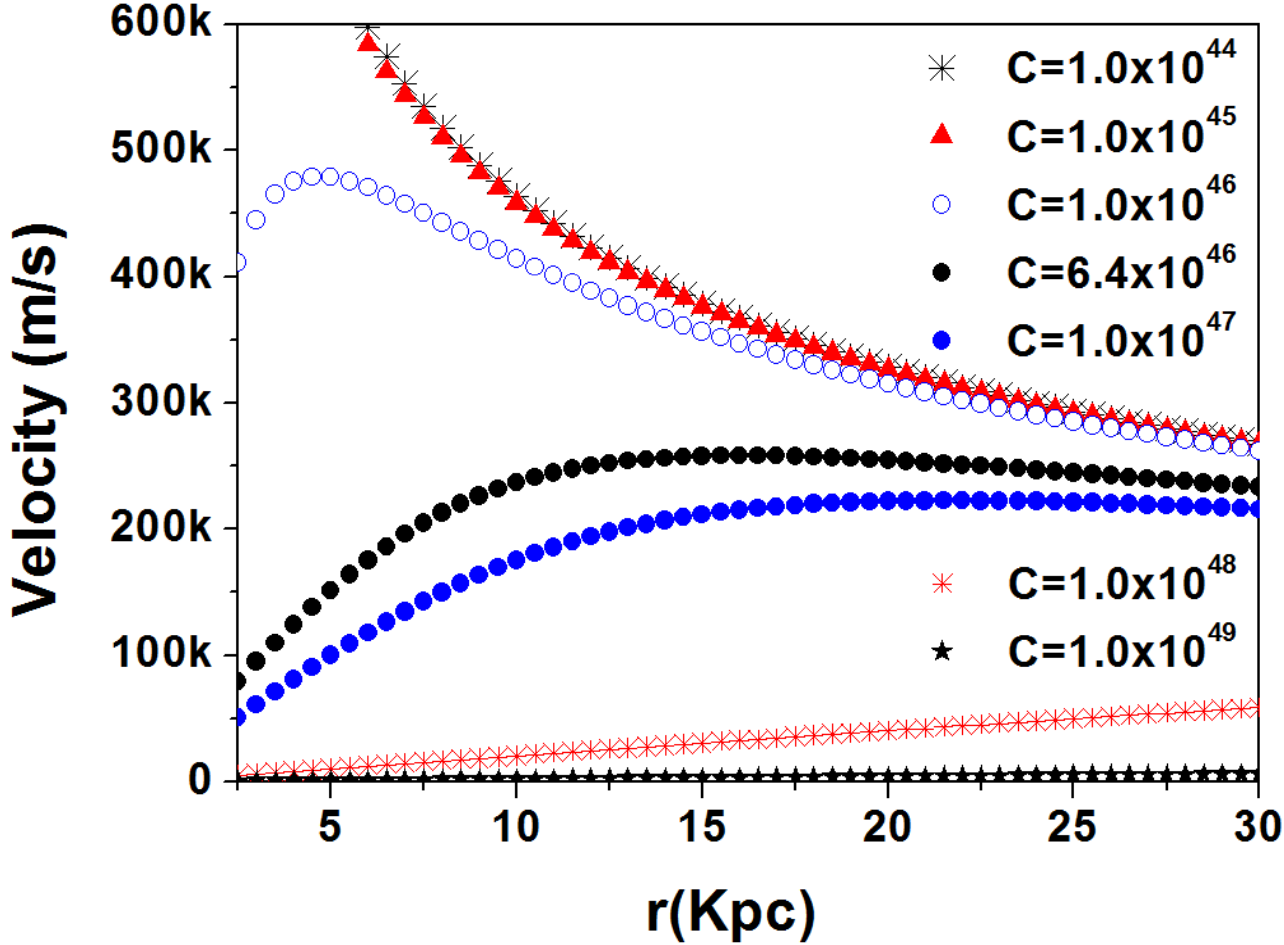

Figure 4

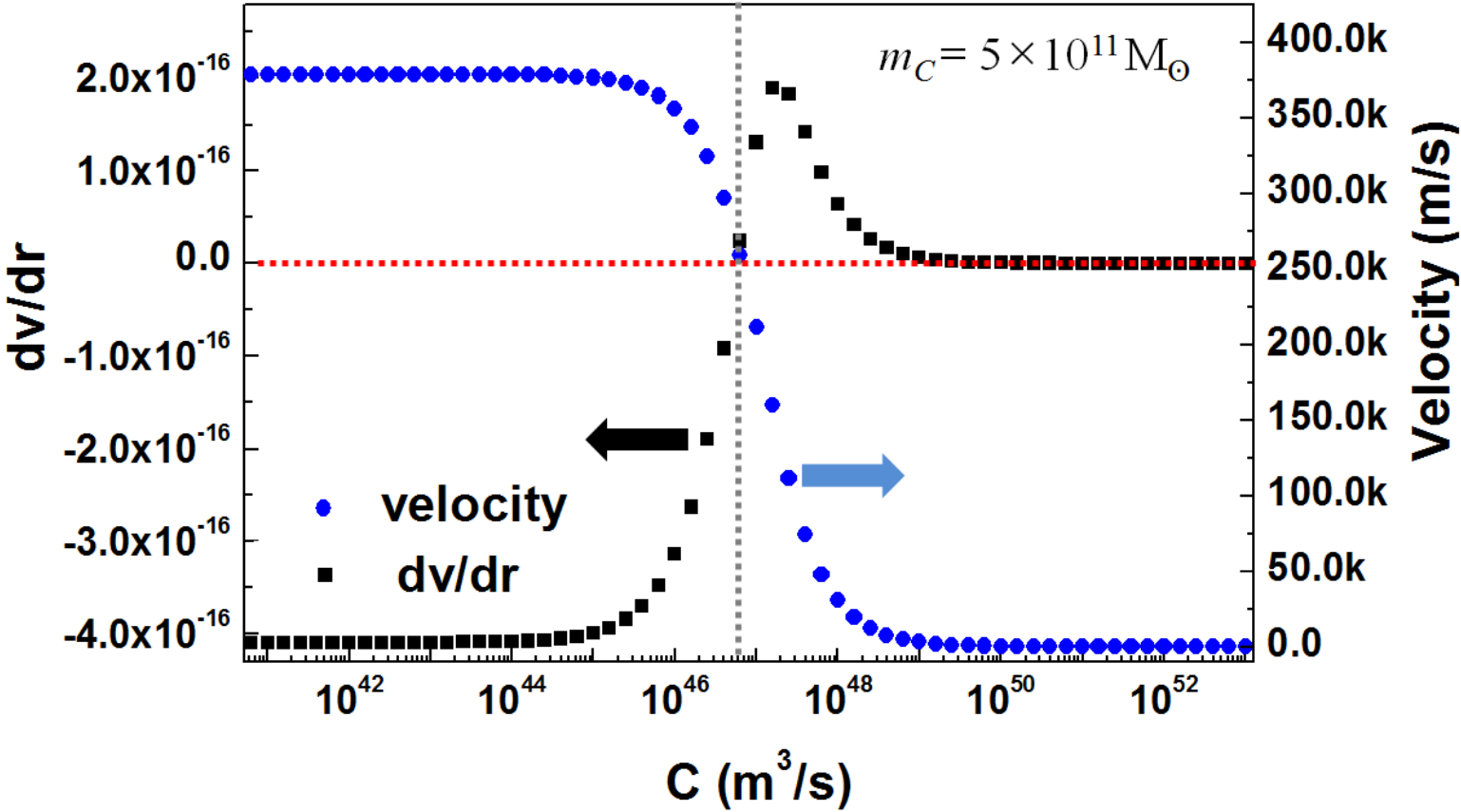

Figure 5

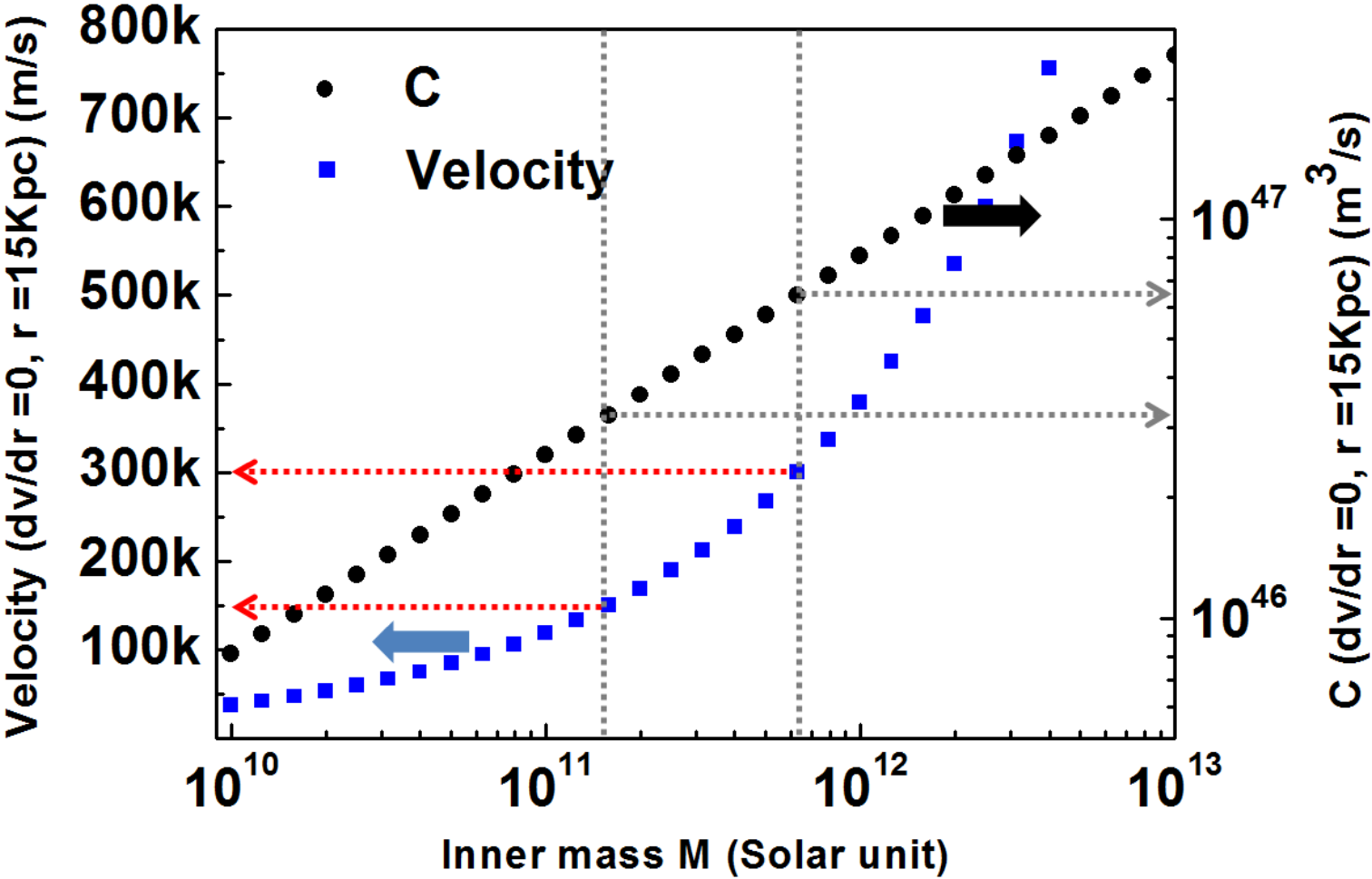